\documentstyle[aps]{revtex}
\begin{document}
\newcommand{\beq}{\begin{equation}}
\newcommand{\eeq}{\end{equation}}
\newcommand{\beqn}{\begin{eqnarray}}
\newcommand{\eeqn}{\end{eqnarray}}
\newcommand{\bmath}{\begin{mathletters}}
\newcommand{\emath}{\end{mathletters}}
\twocolumn[\hsize\textwidth\columnwidth\hsize\csname @twocolumnfalse\endcsname 
\title{ Quantum Monte Carlo and exact diagonalization study of a dynamic Hubbard model}
\author{J. E. Hirsch }
\address{Department of Physics, University of California, San Diego\\
La Jolla, CA 92093-0319}

\date{December 31, 2001} 
\maketitle 
\begin{abstract} 
A one-dimensional model of electrons locally coupled to spin-1/2 degrees of freedom
is studied by numerical techniques. The model is one in the class of $dynamic$
$Hubbard$ $models$ that describe the relaxation of an atomic orbital upon double
electron occupancy due to electron-electron interactions. We study the parameter
regime where pairing occurs in this model by exact diagonalization of small
clusters. World line quantum Monte Carlo simulations support the results of exact
diagonalization for larger systems and show that kinetic energy is lowered
when pairing occurs. The qualitative physics of this model and
others in its class, obtained through approximate analytic calculations, is that
superconductivity  occurs through hole undressing even in parameter
regimes where the effective on-site interaction is strongly repulsive. 
Our numerical results confirm the expected qualitative behavior, and 
show that pairing will occur in a substantially larger parameter
regime than predicted by the approximate low energy effective Hamiltonian.
\end{abstract}

\vskip2pc]
 
\section{Introduction}
Dynamic Hubbard models have been recently introduced as a new class of model 
Hamiltonians to describe the relaxation of atomic orbitals when electrons
are added to orbitals already occupied by other electrons\cite{hole1,dynhub,hole2}. 
This process, originating in the strong on-site repulsion between electrons
in the same atomic orbital, is not described by the conventional Hubbard 
model\cite{hub}. In dynamic Hubbard models this physics is represented either by 
introducing auxiliary spin\cite{spin} or oscillator\cite{polaron} degrees of freedom, or by adding
a second electronic orbital to the site Hilbert space\cite{electron}, with suitable
interaction parameters. As a consequence, the on-site Hubbard repulsion
becomes a dynamical variable and can take a range of values rather than
a single fixed value as in the static (conventional) Hubbard model.
It has been proposed that this physics is
ubiquitous to electrons in atoms, molecules and solids\cite{hole1,dynhub,hole2,hole3}, 
and that it is relevant 
to the understanding of superconductivity in nature\cite{hole4}.

While a vast amount of work has been performed over the years on the
conventional Hubbard model\cite{hub2,hub3}, very little work has been done so far on
dynamic Hubbard models. It is known \cite{hole5,polaron,electron} that in the strong coupling 
anti-adiabatic limit these models map onto the Hubbard model with
correlated hopping, i.e. a Hubbard model where the electronic
hopping amplitude depends on the occupation of the two sites involved 
in the hopping process. This model is known to exhibit superconductivity
when the Fermi level is close to the top of the band, both from mean
field calculations\cite{hole6,micnas,appel}, 
exact diagonalization\cite{aligia,parola,lin}, and other exact 
techniques\cite{zitt,mull}. Furthermore, a variety of observable properties have been
calculated in this limit such as thermodynamics\cite{hole6,mars}, tunneling\cite{tunn},
optical properties\cite{optical}, pressure dependence\cite{mars}, etc. 
Because superconductivity occurs in the
dilute carrier concentration regime it is believed that these
BCS mean field calculations are reliable\cite{dilute}.

The antiadiabatic limit of these models occurs when the frequency of the 
associated boson degree of freedom, $\omega_0$, is much larger than the
effective hopping amplitude for the electrons (small polaron regime\cite{holst}), where the
boson follows the electronic motion. In that limit the parameter regime where pairing
occurs can be calculated exactly for a dilute concentration of hole carriers\cite{dilute}.
Furthermore, numerical calculations on finite clusters show that the doping
regime where pairing occurs is accurately estimated by BCS theory\cite{lin}. 
For finite frequency $\omega_0$, some numerical results have been reported\cite{spin,hole5}.
However it is generally not known whether finite $\omega_0$ enhances or
reduces the tendency to pairing.

Furthermore, in the antiadiabatic limit the single carriers have large
effective mass, and the effective mass is lowered when carriers pair\cite{bose,london}.
The resulting gain in kinetic energy drives superconductivity\cite{nine}. It is not
known whether this physics exists beyond the antiadiabatic limit.

In this paper we study a particular realization of a dynamic Hubbard model,
with an auxiliary spin degree of freedom, by exact diagonalization of small
clusters and a quantum Monte Carlo method, to shed light onto the properties
of the model away from the antiadiabatic limit. We believe that similar
qualitative behavior may be found in the entire class of dynamic Hubbard
models. Briefly, our results show that the qualitative physics of the
antiadiabatic limit persists for finite $\omega _0$, and that the parameter
regime where pairing occurs can be substantially larger. Even though our
results are for a one-dimensional system, we believe it is likely
that the same occurs in higher dimension.

The model studied here bears some superficial resemblance to electron-boson models that have been
extensively studied in the past such as the Holstein model\cite{holst}. However its physics
is qualitatively different. To illustrate this point we present some numerical results 
 for an electron-hole symmetric model with an auxiliary spin degree of freedom
coupled to the electronic site density. This model is expected to be similar
to the Holstein model, and exhibits qualitatively different physics to the
dynamic Hubbard model.

The paper is organized as follows. Section II defines the model and discusses
its properties in the antiadiabatic limit. In Sect. III we present results for
the effective interaction and kinetic energy from diagonalization of small
clusters, and Sect. IV discusses results of world line quantum Monte Carlo
simulations. In Sect. V we present and discuss results for the
electron-hole symmetric Holstein-like model. Sect. VI  discusses the relation between 
the dynamic Hubbard model studied here for a 
site and a real atom.  We conclude in Sect. VII with a summary of our results and
a discussion of the many open questions in this area.

\section{Dynamic Hubbard model with spin-1/2 degree of freedom}
The essence of dynamic Hubbard models is electron-hole symmetry breaking
at the local (1-site) level, so that the dressing of a hole is larger than
the dressing of an electron\cite{hole2}. This physics originates in the dynamic lowering of the
on-site repulsion $U$ when a second electron is added to an atomic orbital,
due to rearrangement of the first electron, and is a ubiquitous phenomenon
in atoms\cite{dynhub}.
There are a variety of  dynamic Hubbard models that can be constructed with
an auxiliary spin-1/2 degree of freedom\cite{hole1,hole2}. Here we consider the site 
Hamiltonian $for$ $electrons$
\bmath
\beq
H_i=\omega_0 \sigma_x^i + g \omega_0 \sigma_z^i
+[U-2g\omega_0\sigma_z^i]n_{i\uparrow}n_{i\downarrow}
\eeq
Hence, for zero and one electrons (one and two holes) at the site, the
site Hamiltonian  is
\beq
H_i(n_i<2)=\omega_0 \sigma_x^i + g \omega_0 \sigma_z^i
\eeq
and for two electrons (zero holes) at the site it is (spin part only)
\beq
H_i(n_i=2)=\omega_0 \sigma_x^i - g \omega_0 \sigma_z^i 
\eeq
\emath
Eq. (1a)  can  be written in hole representation as
\beqn
H_i&=&\omega_0 \sigma_x^i+ g \omega_0 [2(n_{i\uparrow}+n_{i\downarrow})-1]\sigma_z^i
\nonumber \\
& &+[U-2g\omega_0\sigma_z^i]n_{i\uparrow}n_{i\downarrow}
\eeqn
(omitting a chemical potential term), and the lattice Hamiltonian is
\beq
H=\sum_i H_i -t\sum_{i,\sigma} [c_{i\sigma}^\dagger c_{i+1,\sigma} + h.c.]
\eeq
in either electron or hole representation. The electron-hole
transformation is $c_{i\sigma}^\dagger \rightarrow (-1)^i c_{i\sigma}$.

\subsection{Site Hamiltonian}
We will use the Hamiltonian in the hole representation, Eq. (2). The
site eigenstates when there are $n$ holes at the site are, in terms
of the spin-1/2 $\sigma_z$ eigenstates $|+>$, $|->$
\bmath
\beq
|n>=u(n)|+>+v(n)|->
\eeq
\beq
|\bar{n}>=v(n)|+>-u(n)|->
\eeq
\emath
with eigenvalues (excluding the $\sigma$-independent term in $H_i$)
\beq
\epsilon(n)=-\epsilon(\bar{n})=-\omega_0\sqrt{1+g^2}
\eeq
and
\bmath
\beq
u^2(0)=\frac{1}{2}(1+\frac{g}{\sqrt{1+g^2}})
\eeq
\beq
v^2(0)=\frac{1}{2}(1-\frac{g}{\sqrt{1+g^2}})
\eeq
\beq
u(0)v(0)=-\frac{1}{2\sqrt{1+g^2}}
\eeq
\beq
\frac{u(0)}{v(0)}=g-\sqrt{1+g^2}
\eeq
\emath
and
\bmath
\beq
u(1)=u(2)=v(0)
\eeq
\beq
v(1)=v(2)=u(0).
\eeq
\emath
Hence the ground state energy is independent of the electronic site occupation in
this model. The site eigenfunctions depend on $g$ but not on
$\omega_0$, and are the same for site occupation $n=1$ and $n=2$, and
different for $n=0$.  For large $g$ the ground state wavefunctions are almost eigenstates
of $\sigma_z$, with $\sigma_z\sim -1$ for one-hole and two-hole occupation,
very different from the one for zero hole occupation for which $\sigma_z\sim +1$;
while for small $g$ the ground state wavefunction is almost an eigenstate
of $\sigma_x$ ($\sigma_x\sim -1$) and similar for the different hole occupations.
The site eigenvalues depend on both $g$ and $\omega_0$.

The on-site repulsion between two holes (or two electrons)
at the same site depends on the state of the spin degree of freedom and
can range between $U+2g\omega_0$ and $U-2g\omega_0$. The effective on-site
repulsion however, since the ground state energy Eq. (5) is independent of
occupation, is simply $U$ . Note that our notation here is different from that of
Ref. \cite{dynhub}, where $U$ denoted the $bare$ on-site repulsion; here,
the bare on-site repulsion, which is the on-site repulsion if the
background degree of freedom is not allowed to relax upon
double occupancy, is 
\beq
U_{bare}=U+\frac{2g^2\omega_0}{\sqrt{1+g^2}}  ,
\eeq
or $U_{bare}\sim U+2g\omega_0$ for large $g$. Finally, the overlap matrix elements 
between the ground state wavefunctions for the various hole occupations are
\bmath
\beq
<0|1>=2u(0)v(0)=\frac{1}{\sqrt{1+g^2}}\equiv S
\eeq
\beq
<1|2>=1
\eeq
\emath

\subsection{Effective low energy Hamiltonian}
The effective hopping amplitude for a hole between neighboring sites
if the spin degree of freedom makes a ground-state to ground-state
(diagonal) transition is
\bmath
\beq
t_2=|<0|1>|^2 t=S^2t=\frac{t}{1+g^2}
\eeq
if there are no other holes in the two sites involved in the hopping
process. Instead, if there are either one or two other holes of opposite
spins the hopping amplitudes are
\beq
t_1=|<0|1><1|2>|=S t
\eeq
\beq
t_0=|<1|2>|^2t=t
\eeq
\emath
respectively. The low energy effective Hamiltonian for holes in the small polaron
regime is then
\bmath
\beq
H_{eff}=-\sum_{i,\sigma}t_{i,i+1}^\sigma (c_{i\sigma}^\dagger c_{i+1,\sigma}+h.c.)
+U\sum_i n_{i \uparrow}n_{i \downarrow}
\eeq
\beq
t_{ij}^\sigma=t[S^2+S(1-S)(n_{i,-\sigma}+n_{j,-\sigma})
+(1-S)^2 n_{i,-\sigma}n_{j,-\sigma}]
\eeq
\emath
In the regime of low hole concentration the hopping processes where 
more than two holes are in the sites involved can be neglected, and the
effective Hamiltonian is the Hubbard model with correlated hopping\cite{hole6}
\bmath
\beqn
H_{eff}&=&-\sum_{i,\sigma}[t_2+\Delta t(n_{i,-\sigma}+n_{i+1,-\sigma})]
(c_{i\sigma}^\dagger c_{i+1\sigma}+h.c.) \nonumber \\
& &+U\sum_i n_{i \uparrow}n_{i \downarrow}
\eeqn
\beq
\Delta t =t_1-t_2=tS(1-S)
\eeq
\emath

The binding energy of the polaron is obtained from the difference
of $\epsilon(0)$ and the expectation value of $H_i$ if the spin
doesn't adjust to the presence of a carrier, and yields
\beq
\epsilon_p=\frac{2\omega_0 g^2}{\sqrt{1+g^2}}
\eeq
The criterion for small polaron formation is that the energy of the polaron
is smaller than that of a carrier that moves without changing the spin
background:
\beq
z(t-t_2)<\epsilon_p
\eeq
with $z$ the number of nearest neighbors to a site ($z=2$ in one dimension).
Eq. (14) yields
\beq
t<\frac{2\omega_0}{z} \sqrt{1+g^2}
\eeq
as the condition for polaron formation. For the hopping of a single polaron,
the antiadiabatic limit is valid if the polaron hopping amplitude $t_2$ is smaller
than the spacing between site energy levels, hence 
\beq  t<\omega_0(1+g^2)^{3/2}   \eeq
which is always satisfied if the condition Eq. (15) is satisfied.

The condition Eq. (15) indicates that the small polaron regime
will occur when $either$ $\omega_0$ is large $or$ the coupling $g$ is large.
However, the condition Eq. (15) is $not$ sufficient for the effective
Hamiltonian Eq. (11) or (12) to be accurate in the presence of more than one carrier.
Virtual transitions of a hole to a nearest neighbor site occupied by another 
hole yield a contribution to the effective interaction between holes with
an amplitude of the form
\beq
\frac{|<1|\bar{0}>|^2}{\epsilon(\bar{0})+\epsilon(2)-2\epsilon(1)+U}
=\frac{g^2}{1+g^2}\frac{1}{U+2\omega_0\sqrt{1+g^2}}
\eeq
from 'vertical' transitions, which can be much larger than the second
order contribution from the effective Hamiltonian Eq. (11) that describes 
only diagonal transitions:
\beq
\frac{|<1|0>|^2}{\epsilon(0)+\epsilon(2)-2\epsilon(1)+U}=
\frac{1}{(1+g^2)U}
\eeq
These contributions from vertical transitions can be neglected if Eq. (17) is
smaller than Eq. (18), which yields the condition\
\bmath
\beq
\omega_0\sqrt{1+g^2}>U\frac{(g^2-1)}{2}
\eeq
or, for large $g$
\beq
\omega_0 > \frac{Ug}{2} .
\eeq
\emath
Only when $both$ conditions Eq. (15) and Eq. (19) hold can the effective
low energy Hamiltonian Eq. (11) or (12) be expected to be accurate. In particular,
for large $g$ and small $\omega_0$ Eq. (15) may hold and Eq. (19) may not.
In that case, the effective Hamiltonian Eq. (11) can be expected to $underestimate$
the tendency to pairing due to its complete neglect of the site
excited states. The antiadiabatic limit where the effective Hamiltonian
Eq. (11) is valid hence occurs for $\omega _0 \rightarrow \infty$ for
fixed $g$ but not for $g\rightarrow \infty$ for fixed $\omega_0$.
As $g$ increases it is seen from the condition Eq. (19) that the 
antiadiabatic limit will be attained for larger $\omega_0$.

\subsection{Pairing condition and effective mass in antiadiabatic limit}
The condition on the parameters of the Hamiltonian Eq. (12) to yield pairing
of two holes in a full band is (in one and two dimensions)\cite{dilute}
\beq
\frac{\Delta t}{t_2}>\sqrt{1+\frac{U}{D_h}}-1
\eeq
with $D_h=2zt_2$ the single carrier renormalized bandwidth. This is
also the condition for superconductivity within BCS theory in the dilute
limit in any dimension\cite{hole6}. Using Eqs. (12b) and (10a) it translates into
\beq
\frac{U}{D_h}\leq g^2
\eeq
which shows that for $g>1$ pairing will occur even if the on-site repulsion
is larger than the effective bandwidth. Eq. (21) can also be written as
\beq
\frac{U}{D}\leq \frac{g^2}{1+g^2}
\eeq
with $D=2zt$ the unrenormalized bandwidth.

The polaron hopping amplitude increases as the hole filling of the band
increases, according to 
\beq
t(n_h)=t_2+n_h\Delta t
\eeq
with $n_h$ the average number of holes per site ($0\leq n_h \leq 2$), and
correspondingly the bandwidth increases
\beq
D(n_h)=D_h(1+n_h\frac{\Delta t}{t_2})
\eeq
from $D(n_h=0)=D_h$ to $D(n_h=2)=D$. The polaron effective mass correspondingly
decreases as the number of holes increases
\beq
m^*(n_h)=\frac{\hbar ^2}{2t(n_h)a^2}
\eeq
with $a$ the lattice spacing.

When two holes bind in a pair, the pair hopping amplitude $t_p$ in 
the dilute hole concentration regime is 
found to be always larger than $1/2$ the single particle hopping amplitude
\beq
t_p>t_2/2
\eeq
i.e. the pair effective mass is smaller than the sum of the effective 
masses of its constituents\cite{london}. This is opposite to what happens in other
models such as the attractive Hubbard model. Expressions for the
pair mobility $t_p$ are given in refs\cite{bose,london}. The pair
mobility is defined in terms of the energy dispersion relation
for a pair of center of mass momentum q
\beq
E(q)=E_0+t_p q^2
\eeq
and can be obtained by calculating the London penetration depth in
the dilute limit. The kinetic energy per two holes in the dilute limit when there
is no pairing is
\bmath
\beq
<T_s>=-4t_2
\eeq
and when there is pairing the kinetic energy per pair is
\beq
<T_p>=-8t_p 
\eeq
\emath
for a one-dimensional chain.

\section{Exact diagonalization results}
The Hamiltonian of interest has 8 states per site, so that clusters of
up to 8 sites could be studied with current computer capabilities. In this
initial study we restrict ourselves to 2 and 4 sites only. The results
are qualitatively similar and we expect similar qualitative results for larger
clusters, although quantititative differences may be expected for
weak coupling. We compute the effective interaction for two holes in a cluster
from the usual formula
\beq
U_{eff}=2E_0(1)-E_0(0)-E_0(2)
\eeq
with $E_0(n_h)$ the ground state energy for $n_h$ holes;
$U_{eff}<0$ signals a tendency to pairing and superconductivity. For the
$N=2$ cluster the effective interaction in the antiadiabatic limit is
\bmath
\beq
U_{eff}=\frac{U}{2}-\sqrt{(\frac{U}{2})^2+4t_1^2}-2t_2
\eeq
\beq
t_1=t_2+\Delta t=tS
\eeq
\emath
and the condition for pairing ($U_{eff}<0$) is
\beq
\frac{U}{2t}\leq \frac{g^2}{1+g^2}
\eeq
so that in the antiadiabatic limit pairing cannot occur for
$U>2t$ for any value of the coupling parameter $g$. Throughout this
and the following section we will use units so that $t=1$. For the
$N=4$ system, and in fact for any $N\geq 4$ the condition for pairing 
in the antiadiabatic limit is
\beq
\frac{U}{4t}\leq \frac{g^2}{1+g^2}
\eeq
so that pairing will not occur for $U>4$ for any $g$ in the limit
$\omega_0\rightarrow \infty$.

\subsection{Results for effective interaction}
Figure 1 shows the effective interaction for the $N=4$ cluster
as function of coupling constant $g$, for various values of the
on-site repulsion $U$ and two values of the frequency $\omega_0$, together
with the results in the antiadiabatic limit. Note that for small
$\omega_0$ (Figure 	1a) the effective interaction is substantially more
attractive than in the antiadiabatic limit. As $\omega_0$ increases
(Fig. 1b) the results approach those of the antiadiabatic
limit, as expected. The behavior of $U_{eff}$ versus $g$ is non-monotonic 
particularly for small values of $U$.

In Figure 2 we show the dependence of the effective interaction on
$\omega_0$ for the $N=4$ cluster for fixed $U=4$ and
various values of $g$ (a), and for fixed $g$ for various values of $U$ (b).
The limiting values for $\omega_0\rightarrow \infty$ are also shown (dashed lines). 
For $U=4$ there is no pairing in the antiadiabatic limit,
while Fig. 2a shows that for finite frequency pairing will occur for $g\geq 2$.
Similarly, Fig. 2b shows that for $g=3$ (corresponding to an effective mass
enhancement $m^*/m=1+g^2=10$) pairing will occur up to at least
$U=6$ at finite $\omega_0$, while in the antiadiabatic limit $U=3.6$ is the 
maximum on-site repulsion that allows pairing for $g=3$ according to
Eq. (32). Note that for larger $g$ the antiadiabatic limit is approached
for larger $\omega_0$, in accordance with the discussion following Eq. (19).

Even a cluster as small as $N=2$ shows behavior representative of larger clusters
and of (we believe) the thermodynamic limit. The reader can easily verify that
the effective interaction for the $N=2$ cluster obtain by exact diagonalization
closely resembles the behavior of the 4-site cluster shown in Figs. 1 and 2. 
In the antiadiabatic limit the effective interaction as function of $g$ is 
monotononically decreasing with $g$ if the condition for pairing Eq. (31) or (32)
is not satisfied, while if it is satisified it has a (negative) minimum for a finite 
$g$ that decreases as 
$U$ decreases below the limits given by Eqs. (31) and (32).

For an infinite chain, the pair
binding energy can be calculated exactly in the antiadiabatic limit\cite{london}.
The appendix of Ref. 27 gives an analytic expression for the pair binding
energy $\epsilon_b$ in one dimension. The quantity $U_{eff}$ defined by
Eq. (29) calculated here should go to $-\epsilon_b$ as the cluster size
increases. Figure 1 also shows results for $-\epsilon_b$ (dotted lines), which go to zero
when the parameters satisfy the equality in condition Eq. (32). The difference between
the dotted and dashed lines gives the magnitude of finite size effects for
the $N=4$ cluster. It can be
seen that the qualitative behavior of $-\epsilon_b$ for the infinite chain
and $U_{eff}$ for the 4-site chain is the same. The effect of finite size
is to give a somewhat larger attraction, however the condition for pair formation
($U_{eff}<0$) is the same for the N=4 cluster and the 
infinite chain in the antiadiabatic limit (the dashed and dotted lines in Figure 1 go to
zero at the same value of $g$).

In Figure 3 we show the phase diagrams for the $N=2$ and $N=4$ clusters
indicating the region where pairing will occur for some finite frequency
in this model. The full lines  show the results in the antiadiabatic
limit, Eqs. (31) and (32). It can be seen that the region of parameter
space where pairing occurs is substantially enlarged for finite frequency.
Figure 4 shows the optimal frequency for pairing at the phase boundary for
pairing , for the $N=2$ and $N=4$ clusters. For large $g$, $\omega_0$ 
increases slowly with $g$ and is between 1 and 2 (in units of $t$). For 
decreasing $g$, $\omega_0$ goes through a minimum and then diverges, in 
accordance with the fact that the phase boundary lines in Fig. 3 merge with
the ones in the antiadiabatic limit as $g\rightarrow 0$.
It should also be noted that for points away from the phase boundary
the optimal frequency that gives maximum attraction can be
considerable smaller than those shown in Fig. 4 (see e.g. Fig. 2a for
$U=4, g=4$ or Fig. 2b for $g=3,U=2$ where the optimal
frequency is $\omega_0\sim 0.5$).

\subsection{Results for kinetic energy}

The condensation energy in this model is known to be provided by lowering
of kinetic energy in the antiadiabatic limit. Exact expressions for the pair
kinetic energy for two bound holes in a one-dimensional chain are given in
Ref. \cite{london}. Figure 5a shows exact results for the pair kinetic
energy versus coupling constant $g$ for the effective Hamiltonian Eq. (12) for 
various values of $U$. Pairing occurs for couplings obeying the condition
Eq. (32), which for $U=0.8, 2,$ and $3.2$ corresponds to
$g=0.5,1$ and $2$ respectively. For  $g$ larger than those values the kinetic energy
is given by the dashed line, lower than the full line which would be
the kinetic energy in the absence of pairing. Note that even though the
kinetic energy of a pair is lower than that of the unbound holes, it
still $decreases$ in magnitude as the couping $g$ increases. Instead, the
kinetic energy $lowering$, i.e. the difference between the kinetic energy of 
the pair and of the unbound holes, is non-monotonic, peaking at an
intermediate $g$, similarly to the pair binding energy given by
$-U_{eff}$.  Figure 5b shows the kinetic energy lowering per pair
\beq
\Delta T=<T_p>-<T_s>
\eeq
and $U_{eff}$ for the infinite chain, which is the negative of the
pair binding energy $\epsilon_b$ calculated in Ref.\cite{london}.
It can be seen that the two quantities follow similar behavior 
with coupling. In fact, their ratio is essentially constant as function 
of $g$ for large $g$, as shown in Fig. 5c. As the pair binding
decreases, either because $U$ increases or $g$ decreases, both $U_{eff}$ and the 
kinetic energy lowering go to zero. However $U_{eff}$ approaches zero
quadratically\cite{dilute} while the kinetic energy lowering approaches 
zero linearly, hence their ratio diverges as the pair binding energy
goes to zero. 

Note that the kinetic energy lowering upon pairing is
always lower than $-\epsilon_b$. This indicates that the potential energy
change is positive, that is, there is a potential energy cost upon pairing,
given by
\beq
\Delta U_{pot}=(<T_s>-<T_p>)-\epsilon_b
\eeq
and the pair binding energy is smaller than would be expected from the
magnitude of kinetic energy lowering. The potential energy cost arises
from the increased effect of the on-site repulsion between members of a pair since
the pair wavefunction has higher probability for site double occupancy. 

In the infinite chain there is a sharp phase transition between the
state where the pair is bound and where it is unbound, indicated by
the points in Figure 5a where the dashed lines join the full line. 
In the finite chain of course there is no 
sharp transition but rather a smooth crossover. Figure 6a shows results
for the kinetic energy for a pair of holes in the 4-site chain in
the antiadiabatic limit compared
to the results for the infinite chain. As the coupling constant increases the
4-site results cross over from the kinetic energy of unbound holes to the
kinetic energy of the paired holes. When $g$ goes to zero the kinetic energy
of two holes in the 4-site chain is slightly higher than the one for two
unbound holes because of the effect of the on-site repulsion $U$; this is
of course a finite size effect, and for larger clusters and a fixed
number of holes  it will
become negligible in the regime where the holes are not bound.
Figure 6b compares the kinetic energy lowering and the pair binding energy
for the 4-site chain in the antiadiabatic limit and the infinite chain
for one case; it can be seen that both quantities follow similar
behavior, and both are larger in magnitude than for the infinite
chain. Note that in the finite chain kinetic energy lowering goes
to zero for a smaller $g$ than where $U_{eff}$ goes to zero; this would also occur in the
infinite chain for finite hole density. 
We conclude from these results that the 4-site cluster is appropriate to learn
about the qualitative behavior of the kinetic energy for the infinite
chain just as well as
it is for the pair binding energy.

Hence we can now learn about the effect of finite frequency on
kinetic energy lowering by studying the 4-site chain. Figure 7a
shows results for a finite small frequency, $\omega_0=0.5$, compared 
to the antiadiabatic limit $\omega_0=\infty$. Just as for the
effective interaction(Figure 1a), the kinetic energy lowering
can be substantially larger for finite frequency than for infinite frequency, and the
largest kinetic energy lowering occurs for larger $g$ for small frequency. Similarly
Figure 7b shows kinetic energy lowering as function of frequency for
fixed $g$. Similarly as the corresponding results for effective interaction
Figure 2b, the kinetic energy lowering is largest in magnitude at a
fairly low frequency and is considerably larger than in the antiadiabatic
limit. The ratio of kinetic energy lowering to effective interaction
for finite frequencies behaves similarly as in the antiadiabatic limit;
this is shown for one case in Figure 7c.

In addition to kinetic energy lowering it is of interest to consider the
effect of finite $\omega_0$ on kinetic energy itself. Figure 8 shows results
for single hole and pair kinetic energy for finite $\omega_0$ as function of
the on-site repulsion $U$, compared with the limiting case $\omega_0=\infty$.
The difference between the dashed and full lines is the kinetic energy
lowering. It can be seen that relatively speaking the kinetic energy 
lowering is largest in the antiadiabatic limit, even though it is larger
in magnitude for finite $\omega_0$.

In summary, we have seen that the effect of finite frequency
 is to enhance  the pair binding energy and the kinetic
energy lowering found in the antiadiabatic limit. From these results we conclude
that the pair
condensation energy in this dynamic Hubbard model also originates
in kinetic energy lowering, i.e. 'undressing'.
As implied by the conductivity sum rule\cite{apparent,color}, lowering of 
kinetic energy
should be accompanied by  transfer of optical spectral weight from high to low frequencies,
as well as by transfer of spectral weight in the single particle spectral
function from high to low frequencies\cite{undr}. The one and two-particle spectral
functions for the model Eq. (3) will be discussed in a separate paper.

\section{Quantum Monte Carlo simulations}

With quantum Monte Carlo (qmc) methods one can study much larger systems
than with exact diagonalization. We use the basis of $\sigma_z$ eigenstates
for the spin degrees of freedom, so that at every time slice $i$ there are
classical spins $\sigma_j(i)$ at every lattice site $j$. The partition
function is
\beqn
Z=Tre^{-\beta H}=& &Tr \prod_{i=1}^L e^{-\Delta \tau H}= \nonumber \\
& &Tr \prod _{i=1}^L \sum_{{\sigma_j(i)}=+/-1} e^{-\Delta \tau H(\sigma_j(i))}
\eeqn
with $\Delta \tau=\beta/L$, $L$ the number of time slices.
There are two basic 
approaches to quantum
Monte Carlo simulations, determinantal\cite{determ} and world-line algorithms\cite{world}. 

\subsection{Determinantal Monte Carlo}

For the
determinantal algorithm one separates kinetic and potential energy
terms in the Hamiltonian into the product of two exponentials, and decouples the 
interaction term by a discrete Hubbard-Stratonovich transformation
introducing auxiliary Ising variables which we call $\mu_j(i)$ here\cite{determ2}:
\bmath
\beqn
& &e^{-\Delta \tau U_j(i) n_{j\uparrow}n_{j\downarrow}}=\\ \nonumber
& &\frac{1}{2}\times 
\sum_{\mu_j(i)=+/-1} e^{\lambda(\sigma_j(i))\mu_j(i) 
(n_{j\uparrow}-n_{j\downarrow})- \frac{\Delta \tau U_j(i)}{2} (n_{j\uparrow}+n_{j\downarrow})}
\eeqn
\beq
cosh \lambda(\sigma_j(i))=e^{\Delta \tau U_j(i)}
\eeq
\beq
U_j(i)\equiv U(\sigma_j(i))=U-2g \omega_0 \sigma_j(i)
\eeq
\emath
In contrast to the ordinary Hubbard model, the parameter $\lambda$ here is
not constant but depends on the local $\sigma$ variable. For the 
transformation Eq. (36) to be valid it is necessary that
$U_j(i)$ is positive for all values of $\sigma_j(i)$,
i.e. $U\geq 2g\omega_0$.
Next one  takes the trace over fermion degrees of freedom analytically to
obtain the fermion determinant, and the Monte Carlo simulation proceeds by
sampling the Ising spin degrees of freedom $\sigma$ and $\mu$ at each space-time
site. Even though negative weights may occur in this formulation, we do not
expect that they will be very significant in the dilute regime of interest for
this model.

The path integral formulation provided by Eq. (35), (36) makes the 
nature of this dynamic Hubbard model particularly apparent. The
Hubbard $U$ here has space-time fluctuations, with possible
values $U_j(i)=U-2g\omega_0$ and $U_j(i)=U+2g\omega_0$. This
fluctuating $U$ corresponds to the different values that the
on-site Coulomb repulsion between 2 electrons will take
depending on the relative state of these electrons, and embodies
the physics of intra-atomic electronic correlation (at least for
non-degenerate atomic orbitals). In a more realistic dynamic Hubbard
model the Hubbard $U$ will take a continuum of different values.
The energy scale that determines the fluctuations in $U$, $\omega_0$ in this
case, is a one-electron energy scale that reflects the cost in one-electron
energy as the electrons sample the various atomic states to reduce the
magnitude of their intra-atomic Coulomb repulsion.

The determinantal algorithm can be used for the lattice problem as well
as for impurity problems, and as part of the dynamical mean field theory
solution of the model. This  will be deferred to future work. Here we
will instead use the world line qmc algorithm.

\subsection{World line Monte Carlo}

The world line Monte Carlo algorithm can be used if the system is
one-dimensional so that  no negative-weight problems arise. The partition
function is written as
\beq
Z=Tr \prod_{i=1}^L e^{-\Delta \tau \sum_j H_j}
e^{-\Delta \tau H_{kin}^e}e^{-\Delta \tau H_{kin}^o}
\eeq
where the kinetic energy part of the Hamiltonian was decoupled in terms involving
even and odd sublattices. The trace in Eq. (37) is performed by
introducing intermediate states in the spin $\sigma_z$ representation
and the fermion occupation number representation in the usual way.
In addition to moving fermion world lines and flipping individual
$\sigma$ spins we also use composite moves consisting in moving a
fermion world line and flipping the spins at the sites
where the fermion occupation is changing. These moves are necessary to
achieve equilibration in the strong coupling regime.

Figure 9 shows typical world line configurations for two holes and the
associated boson field in a strong coupling regime, with $g=3$ and $\omega_0=2$.
We start the holes far from each other, in the first snapshot shown (a) 
after several hundred sweeps they are still far apart and the world lines 
are rather straight, corresponding to large hole effective mass. After
several more hundred sweeps the holes bind in a bipolaron, as seen in
Figure 9b. The bipolaron has a smaller effective mass, as indicated
by the larger transverse motion of the world lines in the time-like
direction. These pictures clearly show that upon pairing the carriers
become more mobile in this model.

The relation between pair formation and increase in pair mobility is shown
even more clearly in Fig. 10. Fig. 10 (a) shows the kinetic energy of 
a pair as a function of Monte Carlo sweeps. Each Monte Carlo 'step'
in this figure gives an
average over 30 consecutive sweeps. It can be seen that after approximately
100 steps the kinetic energy becomes lower. At the same time, as
Fig. 10(b) shows, the average distance between the holes decreases
dramatically as the pair is formed.

When the number of holes is increased in the system it is found in the 
BCS solution in the antiadiabatic limit that the tendency to pairing
decreases, the coherence length of the pairs increases until they
eventually dissociate at a critical hole concentration. Similar behavior
is found in exact diagonalization of finite systems in the
antiadiabatic limit. We find here that similar behavior is seen qualitatively
in Monte Carlo simulations for finite frequency. As an example, Fig. 11 shows
snapshots of configurations for 6 holes in a 20-site system, i.e. hole
concentration $n_h=0.15$. The system is started in a disordered
configuration, after several thousand sweeps it could be seen in
snapshots such as Fig. 11a that three well-defined pairs are formed; however,
continuing the run, configurations like Fig. 11b appear, where pairs
overlap and the distance between members of a pair (i.e. the coherence length)
increases. Continuing this run the pairs dissociate completely, later
they form again. These snapshots suggest that for these parameters the
system is close to the pair unbinding transition (it is not clear on which
side). The dependence of critical hole concentration on frequency $\omega_0$ is
unknown and an interesting subject for further study. The fact that the
effective attraction increases for finite frequency in the exact
diagonalization study suggests that the critical hole concentration
may be larger for finite frequency than in the $\omega_0 \rightarrow \infty$
 limit.

To detect a superconducting transition in Monte Carlo simulations it is
considerably simpler to use a grand canonical ensemble formulation as in the
determinantal Monte Carlo methods; in world line Monte Carlo, measurement of
pairing correlation functions would involve breaking world lines which leads to
large fluctations\cite{world}. We can however get some information on
pair binding with the world line method by consideration of the kinetic energy.
As seen in figure 8, the kinetic energy increases gradually in the 4-site
system as the on-site repulsion increases. Figure 12 shows the behavior of 
kinetic energy from Monte Carlo simulations on lattices of size $N=8$ and $N=12$, 
as well as for the static (conventional) Hubbard model ($g=0$). For the
static Hubbard model there is a small dependence of kinetic energy on $U$,
which is due to finite size effects. Instead, for the dynamic Hubbard model
there is a large increase in kinetic energy as $U$ increases from small
values, due to the progressive unbinding of the pair. For sufficiently
large $U$ the pair unbinds and the dependence of kinetic energy on $U$
is weak as in the static Hubbbard model. For the $N=12$ cases a fairly
sharp kink in the kinetic energy indicates the transition point. 

Similarly the existence of pairing can be seen in density-density correlation
functions. Figure 13 (a) and (b) show on-site and nearest neighbor hole-hole
density correlations for the static Hubbard model and the dynamic cases
of Figure 12. The on-site correlation is much larger in the dynamic
case, and approaches the static case values only for large $U$. 
 The nearest neighbor correlation in the dynamic case is much
larger than in the static case and first increases as $U$ increases,
due to the rapid decrease of the on-site correlation. Note also that for large
$U$ the nearest neighbor correlation is still considerably larger
than in the static model, indicating that when the on-site double occupation
is essentially suppressed, retardation gives rise to an effective
nearest neighbor attraction. This can be easily understood from
second order strong coupling perturbation theory.

Electrons behave very different from holes in this model. In figure 14 we show 
snpshots of hole worldline configurations when the band is almost full with holes,
i.e. almost empty with electrons. This is the mirror image of the case shown
in Fig. 9. Here we start the simulation with two electrons on the same
site, and after some sweeps the electrons separate. Furthermore, in
contrast to Fig. 9, the quasiparticle world lines show much larger fluctations
in the time direction indicating the smaller electron effective mass.
The contrast between Fig. 14 and Fig. 9 clearly displays the intrinsic
electron-hole asymmetry of this dynamic Hubbard model. Kinetic energy in this
case shows almost no dependence on $U$, as expected, in contrast to the case
shown in Fig. 12.

In summary, the results of these Monte Carlo simulations support the picture 
obtained from exact diagonalization of small systems: the dynamic
Hubbard model is an effective way to obtain hole pairing driven by kinetic
energy lowering in repulsive fermion systems.

\section{Comparison with results for a Holstein-like model}

The conventional electron-boson models studied in the past involve
coupling of a boson degree of freedom to the electronic charge density
rather than to the double occupancy. Even though it doesn't necessarily follow,
in their simplest form these models are electron-hole symmetric.
We consider here one such model with site Hamiltonian
\beq
H_i=\omega_0 \sigma_x^i+g\omega_0 \sigma_z^i 
[n_{i\uparrow}+n_{i\downarrow}-1]+Un_{i\uparrow}n_{i\downarrow}
\eeq
as a generic model in that class. This model should be similar to the
Holstein model\cite{holst}, where the spin-1/2 degree of freedom is replaced by
a harmonic oscillator. Diagonalization of the site Hamiltonian yields
eigenvalues
\bmath
\beq
\epsilon(0)=\epsilon(2)=-\epsilon(\bar{0})=-\epsilon(\bar{2})=-\omega_0\sqrt{1+g^2}
\eeq
\beq
\epsilon(1)=-\epsilon(\bar{1})=-\omega_0
\eeq
\emath
and eigenvectors of the form Eq. (4), with
\bmath \beq
u=\frac{1}{\sqrt{2}}\sqrt{1+\frac{g}{\sqrt{1+g^2}}}
\eeq
\beq
v=-\frac{1}{\sqrt{2}}\sqrt{1-\frac{g}{\sqrt{1+g^2}}}
\eeq \emath
and effective on-site interaction
\beq
U_{eff}=U-2\omega_0 (\sqrt{1+g^2}-1) .
\eeq
The overlap matrix element between ground state wave functions is
\beq
S=<0|1>=<1|2>-\frac{1}{2} [\sqrt{1+\frac{g}{\sqrt{1+g^2}}}+\sqrt{1-\frac{g}{\sqrt{1+g^2}}}]
\eeq
and ranges between $1$ for $g=0$ to $1/\sqrt{2}$ for $g\rightarrow \infty$, so that it
never becomes small as in the previous case.

Figure 15 shows results of exact diagonalization for an $N=4$ cluster.
As a function of freqency, the effective interaction becomes less
attractive as $\omega_0$ decreases, in contrast to the behavior found
for the dynamic Hubbard model. Furthermore, the kinetic energy increases
as $\omega_0$ increases and the effective attraction increases, as shown
in Fig. 15b. Hence in this model pairing gives rise to kinetic energy
$increase$, and the pair condensation energy originates in 
the larger potential energy decrease, which
is precisely opposite to the behavior in the dynamic Hubbard model.

Figure 16 shows typical world line configurations for two holes in this model.
Similarly to Fig. 9, we start the holes far from each other, in the first
snapshot they are still separate and after several sweeps a pair is formed. Here the
world lines for the pair are rather straight, corresponding to large 
effective mass, while the single holes exhibit larger transverse motion
of the world lines indicating lighter quasiparticles. This is qualitatively
different to the behavior in the dynamic Hubbard model (fig. 9), where the
carriers became lighter when they paired.

The relation between pair formation and pair mobility is also shown in Figure 17,
to be compared with Figure 10 for the dynamic Hubbard model. Initially the hole-hole
distance is large and the kinetic energy is low. When the pairs form after approximately 100 
Monte Carlo 'steps' the hole-hole distance decreases drastically and the
kinetic energy increases, again in qualitative contrast with the
behavior found in the dynamic Hubbard model.

Finally, Figure 18 shows the behavior of the average kinetic energy and of
density-density correlations versus on-site repulsion $U$ for the 
Holstein-like model. As indicated by Fig. 15,
the effective interaction becomes attractive for $g=3$ and $\omega_0=1$ when
$U\sim 4$. This is confirmed by the results of Fig. 18. The on-site
density-density correlation increases sharply as $U$ is decreased below $4$,
indicating pair formation. At the same time, the kinetic energy increases
sharply when the pair forms, again in stark contrast to the behavior of the
dynamic Hubbard model seen in Fig. 12 and 13. The nearest-neighbor density-density
correlation first increases as $U$ increases from $0$, indicating that the
pair wavefunction evolves from describing on-site pairing to
more extended pairing, and then decreases for larger $U$ as the pair dissociates.

In summary, these results suggest that the conventional electron-hole symmetric models
and dynamic Hubbard models define two rather different 'universality classes'.
Both types of models can describe pairing,  with qualitatively different features.

\section{Relation to real atom}

In the dynamic Hubbard model considered in this paper, the on-site
repulsion takes the values $U-2g\omega_0$ and $U+2g\omega_0$ when
the auxiliary spin at the site points up and down respectively in
a $\sigma_z$ representation. More generally, for the spin in a 
superposition of these states the on-site repulsion will take values
intermediate between these extremes. The 'effective' on-site 
repulsion defined by
\beq
U_{eff}(site)=E(2)+E(0)-2E(1)
\eeq
with $E(n)$ the site energy with $n$ electrons (or holes) is simply $U$.
The reason a fluctuating $U$ is needed to represent a real atom is that
the wavefunction of 2 electrons in an orbital is not simply the
product of the single-electron wavefunctions in the singly occupied
atom\cite{hole2}, but rather a superposition
\beq
\Psi(r_1,r_2)=\sum_{n,m} C_{nm}\varphi_n(r_1)\varphi_m(r_2)
\eeq
where $\{\varphi_n(r) \}$ is a complete set of single-electron
wave functions. The fluctuating values of $U$ can be thought of as the
different values that the electron-electron repulsion will take
for one electron in $\varphi_n(r)$ and the other electron in 
$\varphi_m(r)$, for all $n$, $m$ for which $C_{nm}$ is not zero.
The frequency $\omega_0$ represents the energy scale of 
electronic excitations in the atom, i.e. the eigenenergies of the
wavefunctions $\{\varphi_n(r) \}$ .

More specifically, for the particular case of $1s$ orbitals in
a hydrogenic atom of ionic charge $Z$, the 'bare' on-site repulsion
for two electrons in the $1s$ orbital is
\beq
U_1=17Z eV
\eeq
This corresponds in our model to the on-site repulsion
when the boson is not allowed to relax, Eq. (8), or approximately
$U+2g\omega_0$. In the Hartree approximation, the orbital expands
to $\bar{Z}=Z-5/16$ upon double occupation, and the repulsion
between two electrons in these expanded orbitals is Eq. (46) with 
$Z$ replaced by $\bar{Z}$, i.e.
\beq
U_2=17(Z-\frac{5}{16}) eV=U_1-5.31eV
\eeq
This would roughly correspond to the 'minimum' on-site repulsion in
our model, $U-2g\omega_0$. Finally, the effective on-site repulsion
in the Hartree approximation, taking into account the cost in
single-particle energy upon orbital expansion, is
\beq
U_3=17(Z-\frac{5}{32}) eV=U_1-2.66eV
\eeq
precisely halfway between the values Eq. (46) and (47),
and this would correspond to the effective site $U$ in our model,
which is just $U$, also precisely halfway between the minimum and
maximum values. Summarizing,
\bmath
\beq
U_1=17ZeV=U+2g\omega_0
\eeq
\beq
U_2=U_1-5.31eV=U-2g\omega_0
\eeq
\beq
U_3=U_1-2.66eV=U
\eeq
\emath
so that for this case we have simply
\beq
g\omega_0=1.33 eV
\eeq
independent of $Z$.

The frequency $\omega_0$ is related to excitation energies of the atom,
hence we expect the dependence on ionic charge
\beq
\omega_0=cZ^2
\eeq
which implies from Eq. (50) that $g$ increases as the ionic
charge $Z$ decreases. This is in accordance with the fact that the overlap
\beq
S=<0|1>=\frac{1}{\sqrt{1+g^2}}
\eeq
decreases as $g$ increases; in the Hartree approximation, this overlap corresponds
to the overlap of the expanded single electron orbital with the
non-expanded one and is given by 
\beq
S=\frac{(1-\frac{5}{16Z})^{3/2}}{(1-\frac{5}{32Z})^3}
\eeq
which decreases to zero as $Z\rightarrow 0.3125$.

Strictly speaking our dynamic Hubbard model will be a valid
representation of the real atom only in the parameter regime
where
\beq
U-2g\omega_0 > 0
\eeq
because the atomic Coulomb integral for any two orbitals 
$\varphi_n$, $\varphi_m$ has to be positive. Even with
the constraint Eq. (54) a wide range of parametes in the model
exists where pairing will occur, as can be inferred from the numerical
results in the previous sections. For example, from Fig. 1a we see
that for $\omega_0=0.5, U=8$, pairing occurs for $g>4$; the condition
Eq. (54) is satisfied in this case up to $g=8$, and for $g\sim 4$ the
fluctuations in the on-site $U$ are about $50 \%$. As the frequency
gets smaller, the relative fluctuations in $U$ needed to obtain
pairing decrease. For example, from Figure 2a we find that for
$U=4$, $g=4$, pairing occurs for $\omega_0>0.12$, which corresponds
to fluctuations in $U$ of only $25\%$ (between $U=3$ and $U=5$).
If the $U$ was not fluctuating but fixed, no pairing occurs in 
the model unless $U<0$. In other words, the 'equivalent $U$' in
a model with fixed $U$ is not only smaller than the average $U$ in the
fluctuating case but it is smaller than the smallest value that the
fluctuating $U$ attains in these cases. 

Note that as $Z$ in the atom decreases, the on-site bare $U$ decreases
(Eq. (46)), the parameter $\omega_0$ should decrease according to
Eq. (51), and correspondingly $g$ should increase (Eq. (50). As seen
in Figure 2b, as $U$ decreases a smaller $\omega_0$ is required to
give pairing, and as seen in Figure 2a as $g$ increases also a smaller
$\omega_0$ is needed for pairing. We conclude from our results for the
model system and the relationship with the real atom that smaller values
of $Z$ yield the most favorable conditions for pairing in this 
dynamic Hubbard model.

\section{Discussion}

We have studied numerically some properties of a dynamic Hubbard
model, where the value of the on-site repulsion $U$ depends on the
state of an auxiliary boson degree of freedom. In the model studied
in this paper the boson is a spin-1/2 degree of freedom, with
excitation energy $\omega_0$. It will be of interest to study other similar
models with other boson degrees of freedom such as other versions of the
spin 1/2 model\cite{hole1}, higher spin variables or harmonic 
oscillators, or purely electronic models with more than one orbital
per site\cite{hole2}; we believe the qualitative physics will be
similar. Furthermore, 'extended' dynamic Hubbard models, with more than
on-site interactions, should be interesting to study.

These dynamic Hubbard models map onto the Hubbard model with
correlated hopping in the antiadiabatic limit $\omega_0\rightarrow \infty$, 
which is known to lead to pairing of holes and superconductivity
for sufficiently large coupling constant $g$  ;
the purpose of this paper was to determine whether pairing still
exists for finite $\omega_0$. Furthermore in the antiadiabatic limit
pairing is known to occur through kinetic energy lowering, and we examined
whether the same physics occurs for finite $\omega_0$.

From both the exact diagonalization and the quantum Monte Carlo results
we concluded that the same physics of the antiadiabatic limit
persists for finite, and even small, $\omega_0$. Furthermore, the
parameter regime where pairing occurs is larger for small $\omega_0$
than for $\omega_0\rightarrow\infty$. When $\omega_0$ is small
the kinetic energy is much lower than in the antiadiabatic limit, yet the
magnitude of kinetic energy lowering upon pairing is similar to that in the
antiadiabatic limit.

The $\omega_0\rightarrow\infty$ limit of the model (Hubbard model with
correlated hopping) is useful because its physics is rather transparent and
because it allows for much simpler analytic and numerical treatments.
However, strictly speaking the dynamic Hubbard model considered here
is only a realistic representation of a real system for parameters
where $U-2g\omega_0>0$, which certainly does not hold in the
antiadiabatic limit. Hence it is essential to establish that the properties
of the model for small $\omega_0$ and for $\omega_0\rightarrow\infty$
are similar if one is to use the results obtained from the 
$\omega_0\rightarrow\infty$ limit to understand the properties of a real
system. We found that the model can give rise to pairing even in parameter
ranges where the fluctuating $U$ attains only positive values, which implies
that an 'equivalent' fixed $U$ in those cases would be smaller
than the lower bound of the range within which $U$ fluctuates.

We contrasted the behavior found in the dynamic Hubbard model with that of an
electron-hole symmetric Holstein-like model. In the latter model, which we suggest
is representative of a wide range of model Hamiltonians that have been 
considered in the past to describe superconductivity, the physics found is
qualitatively different: pairing is associated with lowering of potential energy
and increase in kinetic energy, opposite to the behavior found in the
dynamic Hubbard model. We suggest that these two models, each representative
of an entire class of model Hamiltonians, are two different paradigms by which
superconductivity can be achieved\cite{paradigm}.  Whether either or both occur in nature is
an unsettled question. The theory of hole superconductivity proposes that only
the paradigm represented by the dynamic Hubbard model occurs in real 
materials\cite{hole3,hole4}.
  
We also discussed briefly the relation between the dynamic Hubbard model considered here 
and a real hydrogen-like atom for $1s$ electrons in the Hartree approximation. Clearly such 
relation should be qualitatively similar for more accurate representations of 
the two-electron wavefunction such as the Hylleraas wave function\cite{hole2}, 
as well as for electrons in other atomic orbitals. 
We found that a smaller value of the ionic charge $Z$ yields more favorable
conditions for pairing for several different reasons: 1) it leads to smaller on-site
repulsion $U$, 2) it leads to larger coupling $g$, which leads to
larger 'dressing' of quasiparticles in the normal state, and to larger
'undressing', hence larger energy lowering, as quasiparticles pair, and 3) it leads to
smaller frequency scale $\omega_0$, which according to the results of
this paper is favorable to pairing as $U$ becomes smaller and $g$ becomes
larger. In addition, smaller $Z$ is also favorable because it leads to larger
orbital overlap between atoms, hence larger bare hopping $t$, which increases
the overall scale of the pairing interaction calculated in this paper\cite{mars}.
It will be interesting to perform
detailed analysis of the connection between this and other dynamic
Hubbard models and electrons in various orbitals in real atoms.

The results presented here are only a first step in the understanding of
this and other dynamic Hubbard models. Analytically, both strong
and weak coupling expansions should be feasible and of interest.
Powerful numerical techniques that have been extensively used for the
static Hubbard model and other related models such as the Holstein model
can and should be brought to bear on this
class of models. In particular, density matrix renormalization
group\cite{dmrg} and determinantal Monte Carlo methods should allow
for the study of larger systems as a function of model parameters and
hole concentration to determine the range of parameters where hole
pairing occurs. The dynamical mean field method combined with a
Monte Carlo 'impurity' method should be a very fruitful approach to deal
with this class of models\cite{dynmf}. In particular, it will be of great interest
to understand quantitatively the processes of spectral weight transfer
in one- and two-particle Green's functions
that are expected to occur in this class of models upon transition
to the superconducting state\cite{color,undr}, which are of interest in connection
with photoemission\cite{photo} and optical experiments\cite{optical2} in superconducting materials.

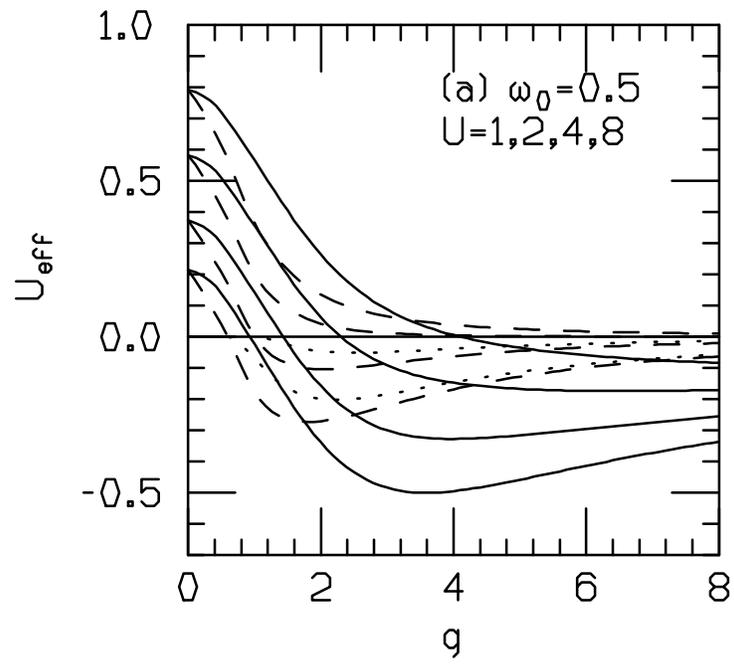
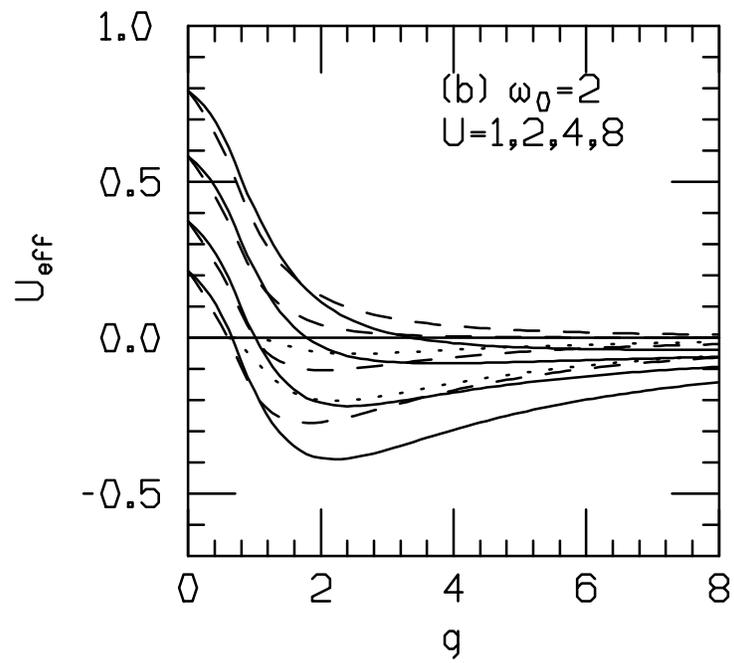
\begin{figure}
\caption { Effective interaction $U_{eff}$ for $N=4$ cluster versus coupling
constant $g$ and various values of the on-site repulsion $U$ for 
(a) $\omega_0=0.5$ and (b) $\omega_0=2$ (full lines). 
The dashed lines and
dotted lines give the results in the $\omega_0 \rightarrow \infty$ limit
for the $N=4$ cluster and for the infinite chain respectively. For 
fixed $g$, increasing $U$ corresponds to increasing value of $U_{eff}$.
}
\label{Fig. 1}
\end{figure}

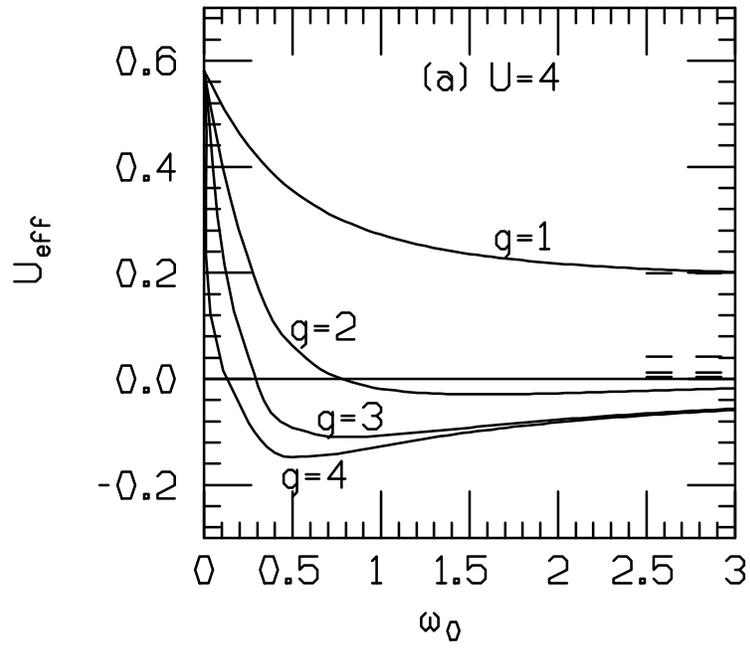
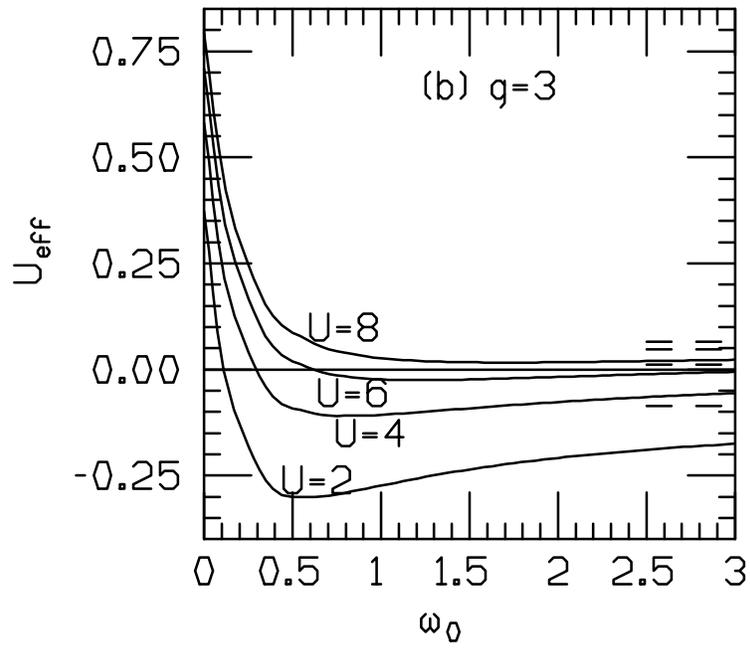
\begin{figure}
\caption { Dependence of $U_{eff}$ on $\omega_0$ for $N=4$ cluster. The 
dashed lines give the limiting values $\omega_0 \rightarrow \infty$.
}
\label{Fig. 2}
\end{figure}

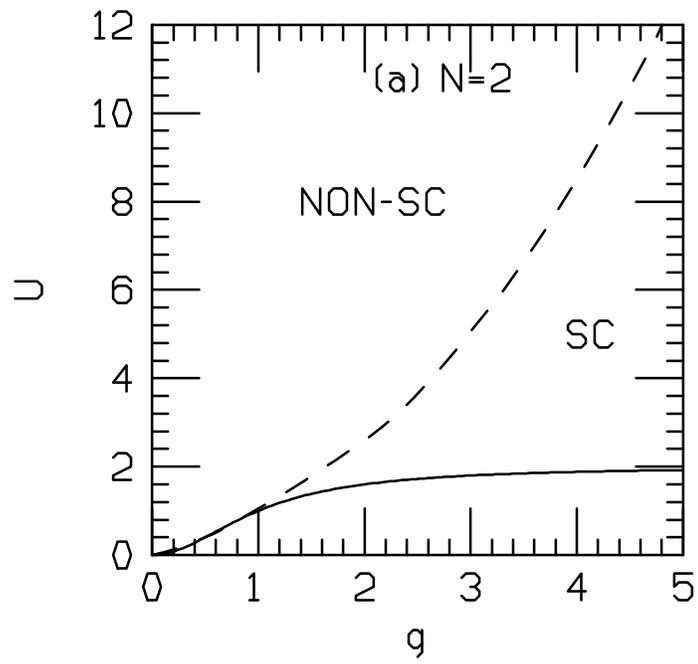
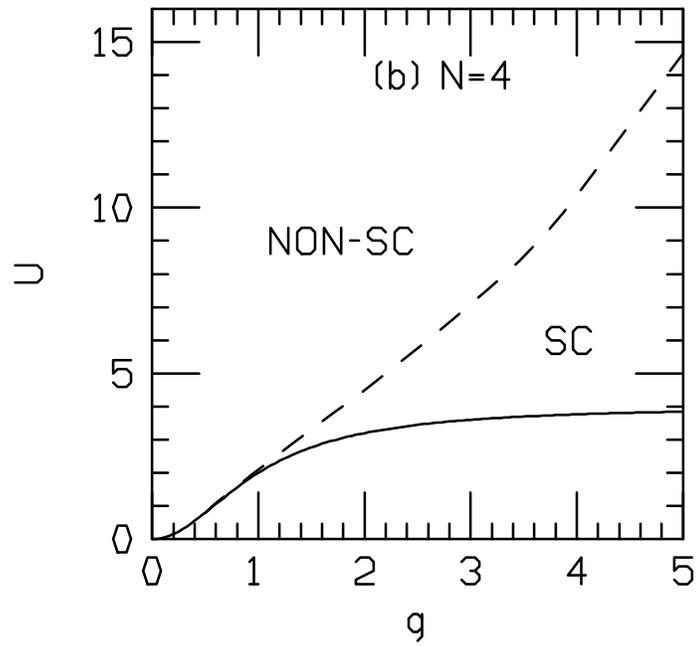
\begin{figure}
\caption { Phase diagram for (a) $N=2$ and (b) $N=4$ clusters. In the
region labeled NON-SC, $U_{eff}>0$ for all values of $\omega_0$; in the
region labeled SC, a range of $\omega_0$ exists where $U_{eff}<0$.
Below the solid line, $U_{eff}<0$ in the antiadiabatic limit
$\omega_0\rightarrow\infty$.
}
\label{Fig. 3}
\end{figure}

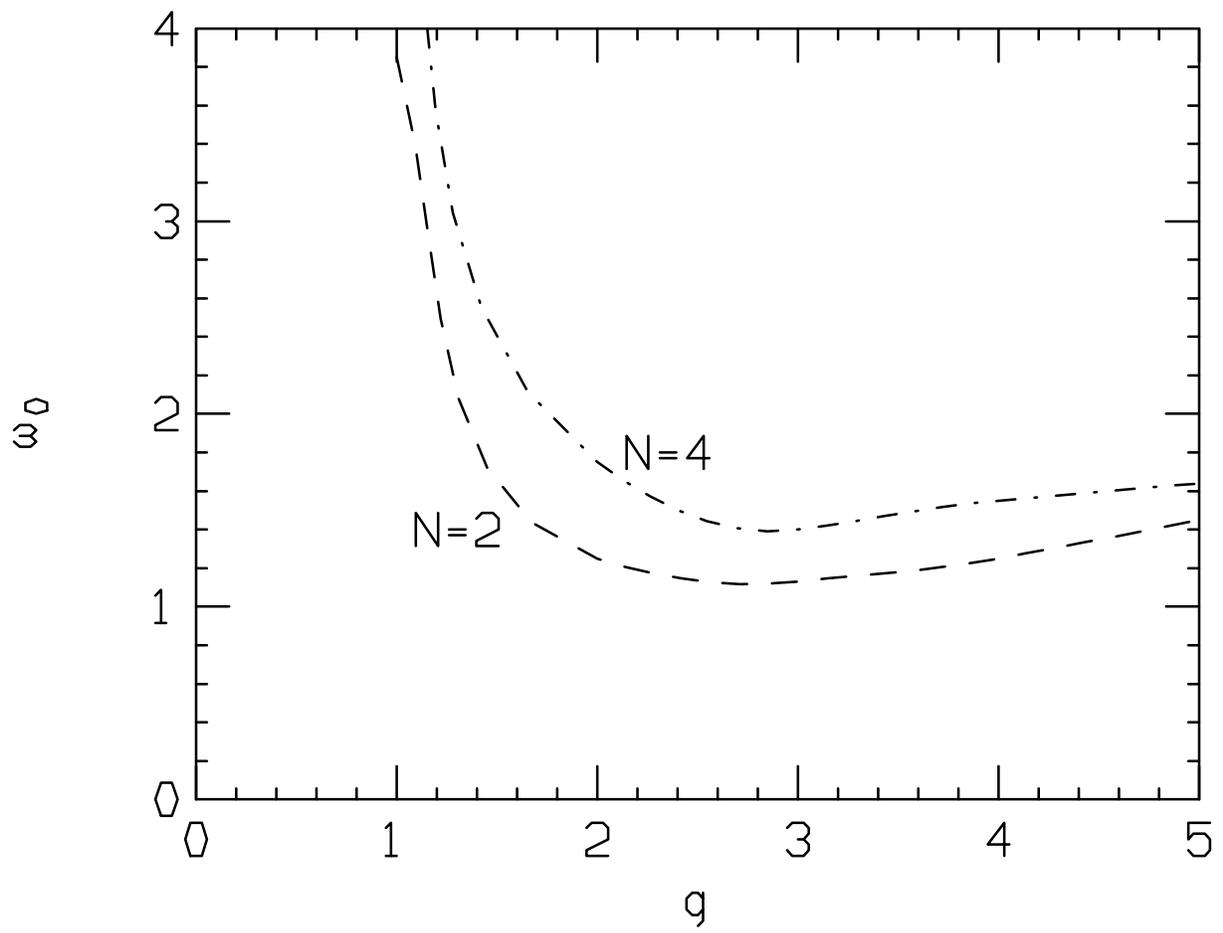
\begin{figure}
\caption { Optimal frequency $\omega_0$ that gives rise to pairing at the phase
boundaries (dashed lines) of figure 3. As $g$ decreases $\omega_0$ increases
and the phase boundary approaches the one in the antiadiabatic limit.
}
\label{Fig. 4}
\end{figure}

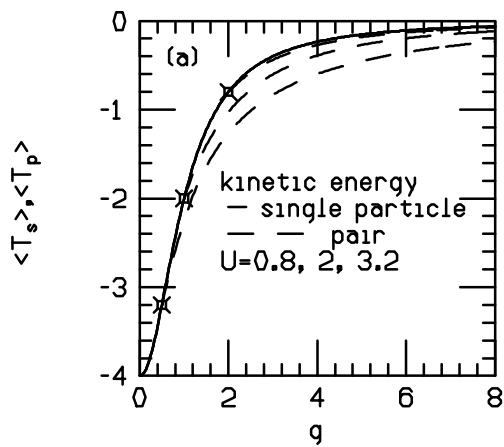
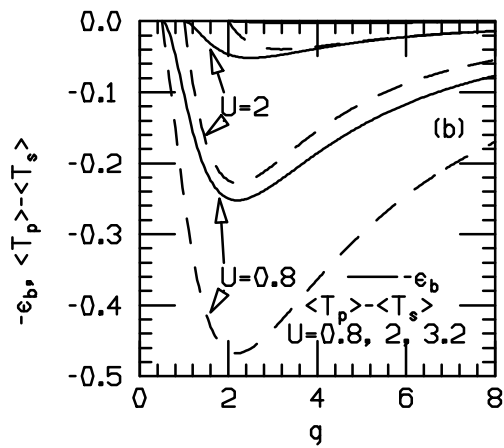
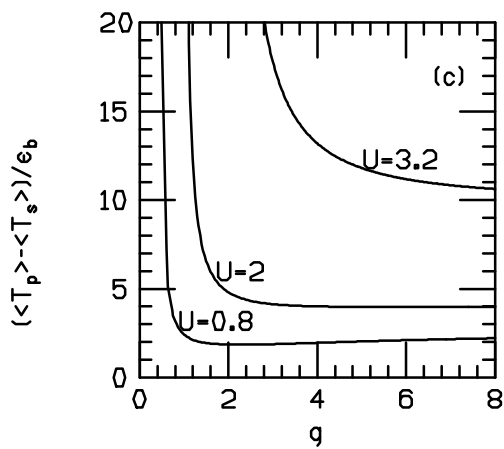
\begin{figure}
\caption { Results for kinetic energy in the infinite chain in the
antiadiabatic limit. (a) Kinetic energy of two unbound holes (full line)
and of a hole pair (dashed lines) versus $g$ for various values of $U$.
As $g$ decreases, the dashed line joins the full line (as indicated
by the symbols) when pairs unbind, at $g=g_c$. $g_c$ is $0.5$, $1$ and $2$
for $U=0.8$, $2$ and $3.2$ respectively. (b) Pair binding energy (full lines)
and kinetic energy lowering (dashed lines) for the infinite chain versus  $g$
for various $U$. At $g_c$, both quantities go to zero. (c) Ratio of
kinetic energy lowering to pair binding energy versus $g$.
}
\label{Fig. 5}
\end{figure}

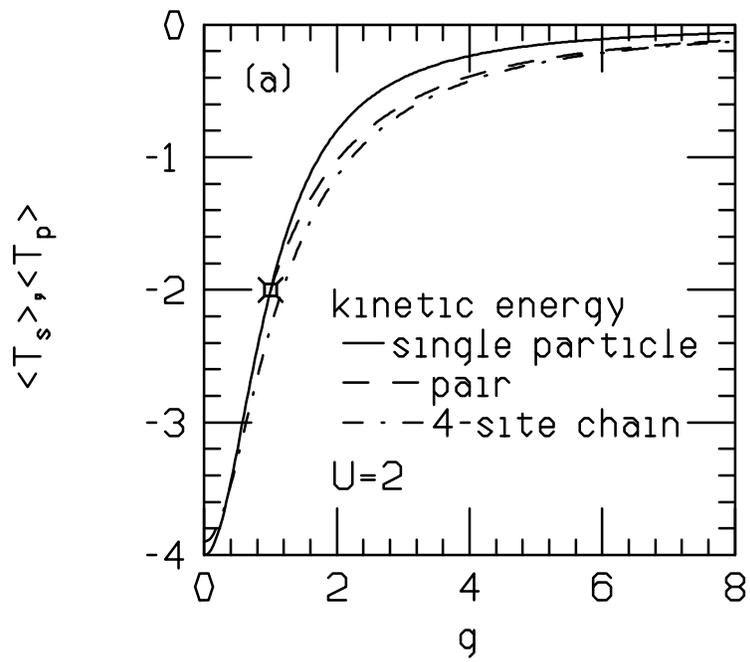
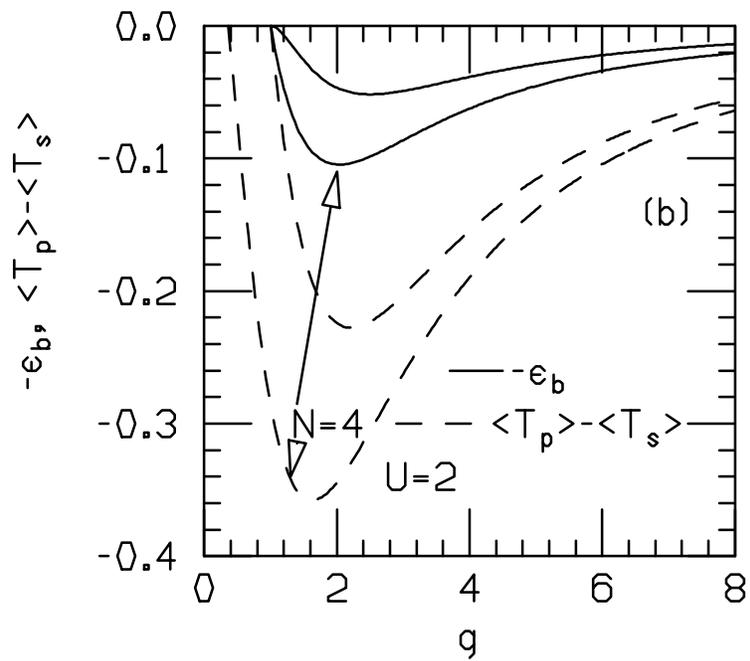
\begin{figure}
\caption { Comparison of results for kinetic energy of $N=4$ chain
and infinite chain in the antiadiabatic limit. In (a), the kinetic 
energy of 2 holes in the $N=4$ cluster (dash-dotted line) joins the
infinite chain kinetic energy of the bound pair for large $g$ (dashed line)
and the kinetic energy of unbound holes for small g (full line). (b) shows
 that finite size effects similarly enhance the magnitude of pair binding
energy and of kinetic energy lowering.
}
\label{Fig. 6}
\end{figure}

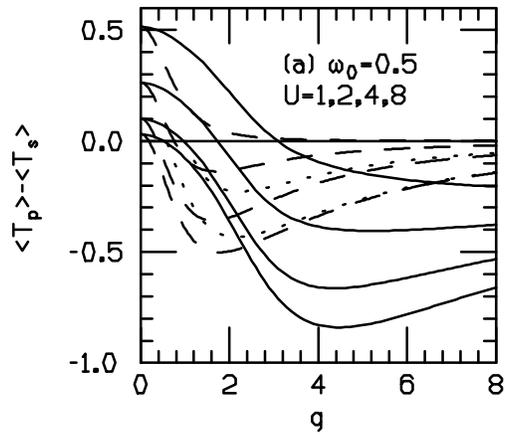
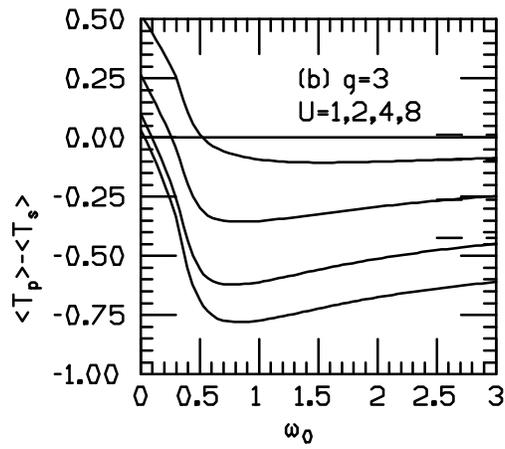
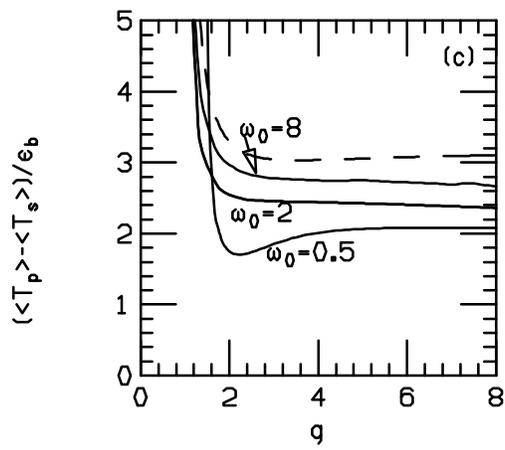
\begin{figure}
\caption { Difference between kinetic energy of a pair and kinetic energy
of two holes in the 4-site chain (a) versus $g$ for fixed $\omega_0$
and (b) versus $\omega_0$ for fixed $g$. The dashed and dotted lines
in (a) give the results in the $\omega_0\rightarrow \infty$ limit for the
$N=4$ cluster and the infinite chain respectively; the dashed lines in
(b) give the results in the $\omega_0\rightarrow \infty$ limit for the
$N=4$ cluster. (c) Ratio of kinetic energy
lowering to pair binding energy in 4-site chain versus $g$ for $U=2$ and
various values of $\omega_0$. The dashed line gives the results for $\omega_0=\infty$.
Note that the kinetic energy lowering upon pairing for small frequency is
considerably larger for finite $\omega_0$ than in the $\omega_0=\infty$
limit. 
}
\label{Fig. 7}
\end{figure}

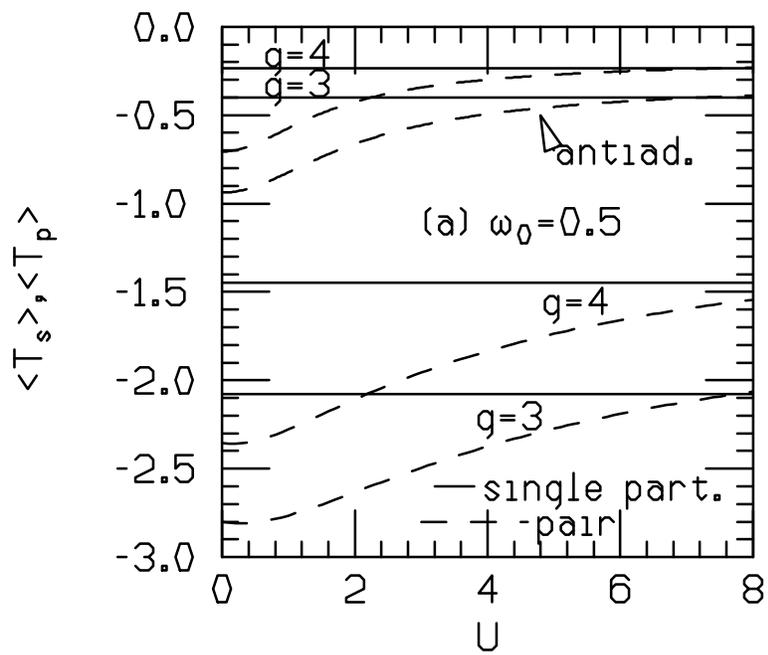
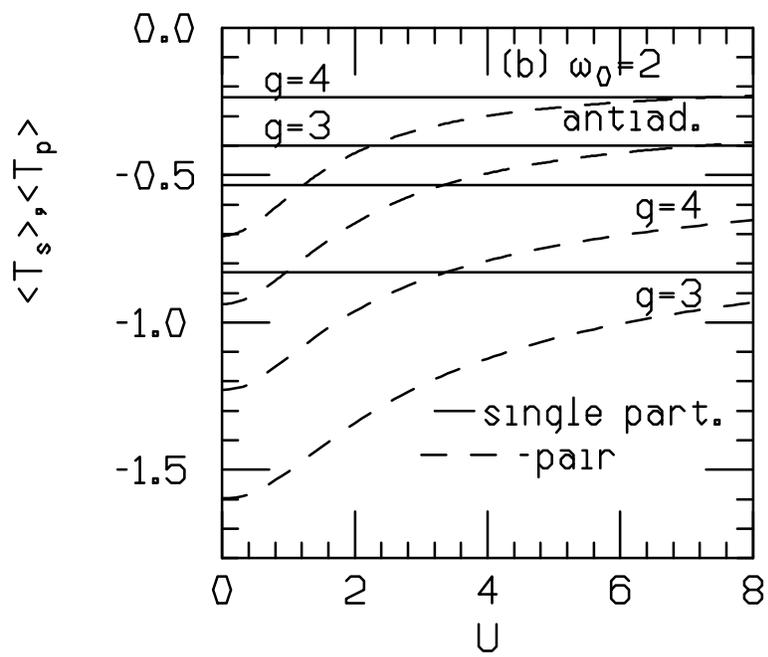
\begin{figure}
\caption { Single particle (full lines) and pair (dashed lines) kinetic energy
versus $U$ for $g=3$ and $g=4$ for (a) $\omega_0=0.5$ and
(b) $\omega_0=2$. The results for $\omega_0\rightarrow \infty$ are also
shown. Both the single particle and the pair kinetic energies are
substantially lower for small $\omega_0$ than for $\omega_0=\infty$.
The transition points where $U_{eff}$ changes sign for $g=3$ and $g=4$ are
respectively $U=3.6$ and $U=3.77$ for $\omega_0=\infty$, 
 $U=6.91$ and $U=10.1$ for $\omega_0=2$ and
 $U=5.57$ and $U=7.77$ for $\omega_0=0.5$. 
}
\label{Fig. 8}
\end{figure}

\begin{figure}
\caption { Snapshots of Monte Carlo configurations for an $N=20$-site
lattice with $L=40$ time slices. $\Delta \tau=0.25$, $g=3$, $\omega_0=2$.
The left-side panel show the hole world lines, the right-hand panel indicates
the boson (spin) configuration. At the sites where the hole occupation is
1 (2), the boson configuration is denoted by $p$ ($b$) if the boson state
is $|->$,  which is the low energy configuration, and by $n$ if it is $|+>$.
In (a), after a few hundred sweeps, the holes are separate and heavy,
in (b), after several more hundred sweeps, the holes are bound and
lighter (world lines fluctuate more in time direction).
}
\label{Fig. 9}
\end{figure}

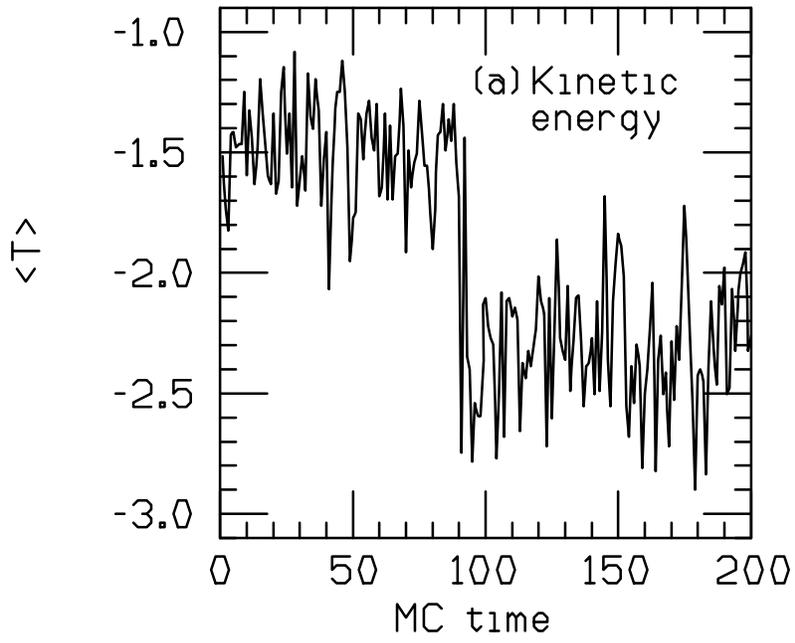
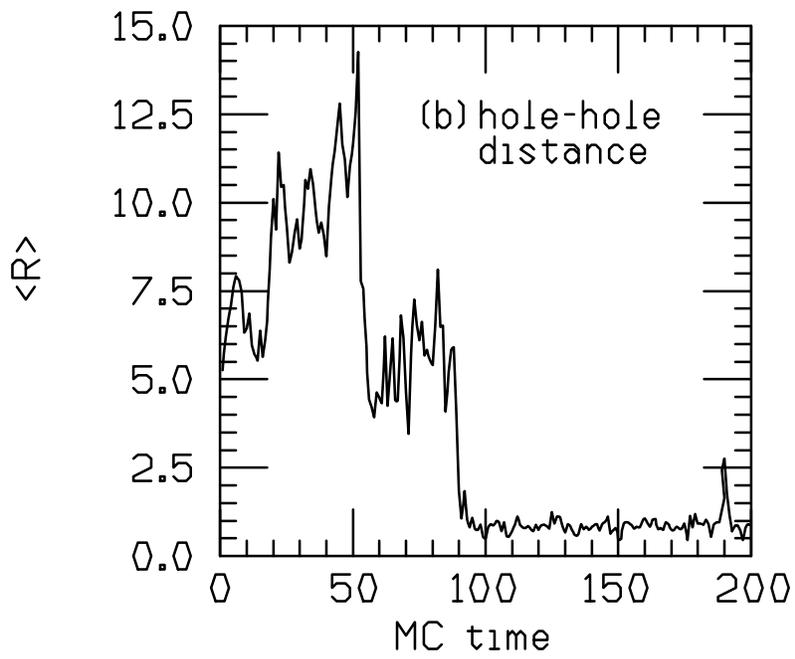
\begin{figure}
\caption {Kinetic energy of two holes (a) and hole-hole distance (b) as function
of 'Monte Carlo time'. The unit of time is 30 Monte Carlo sweeps.  
$N=20$, $L=40$,$\Delta \tau=0.25$, $g=3$, $\omega_0=1$. After about 100
steps the hole-hole distance decreases drastically as the holes become bound (b),
and at the same time the kinetic energy becomes lower (a).
}
\label{Fig. 10}
\end{figure}

\begin{figure}
\caption { Snapshots for 6 holes in a 20-site chain. 
$L=40$, $\Delta \tau=0.25$, $g=3$, $\omega_0=2$. In the first picture (a) three
well-defined pairs are seen, in (b) the pairs overlap and the 
distance between members of a pair ('coherence length') increases.
}
\label{Fig. 11}
\end{figure}

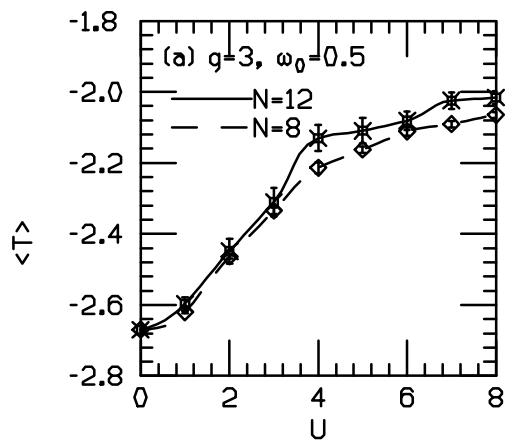
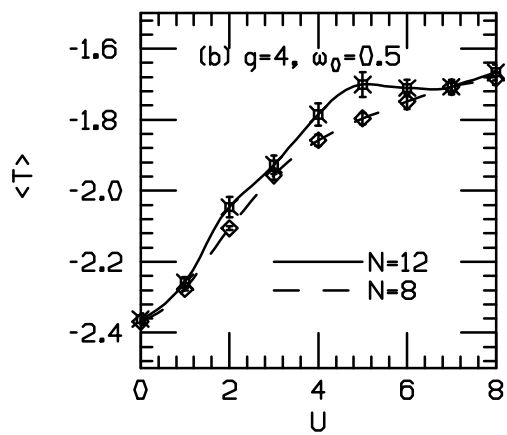
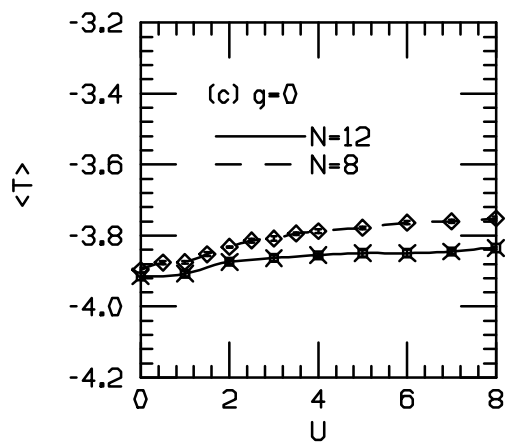
\begin{figure}
\caption { Kinetic energy versus on-site repulsion $U$ from Monte Carlo
simulations of dynamic Hubbard model (a), (b), and of static (conventional)
Hubbard model ($g=0$) (c). 
}
\label{Fig. 12}
\end{figure}

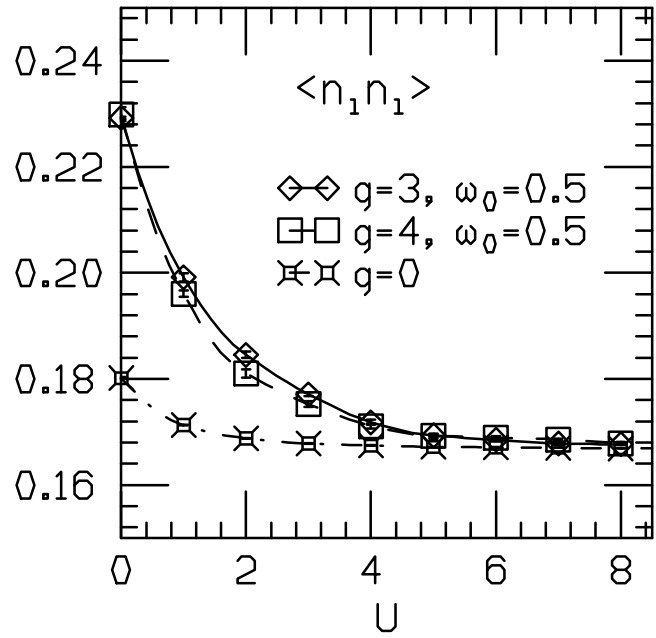
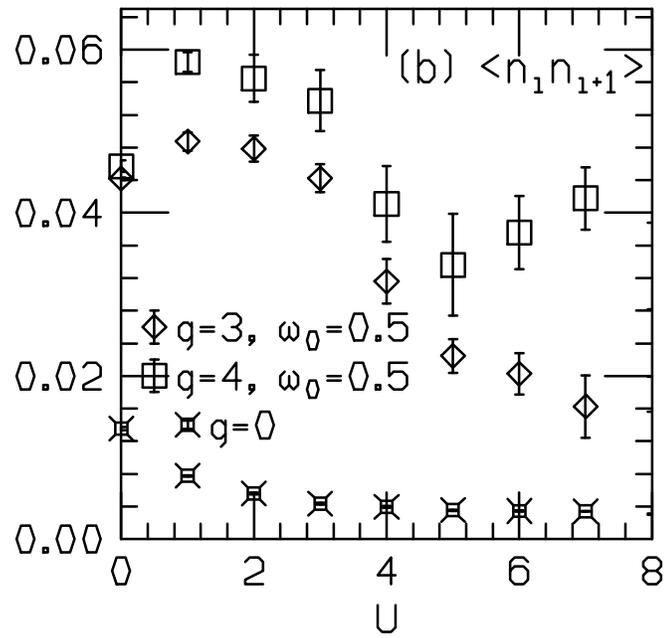
\begin{figure}
\caption { (a) On-site and (b) nearest neighbor density-density correlations
for dynamic and static Hubbard model, $N=12$ sites. 
}
\label{Fig. 13}
\end{figure}

\begin{figure}
\caption { Snapshots of Monte Carlo configurations when the band is almost
full with holes, with only two electrons. The convention and parameters are the same as 
in Fig. 9. The simulation is started with the two electrons on the same
site (top panel), after a few sweeps (bottom panel) the electrons separate
and are light, as illustrated by the large fluctuations of the world lines
in the time direction.
}
\label{Fig. 14}
\end{figure}

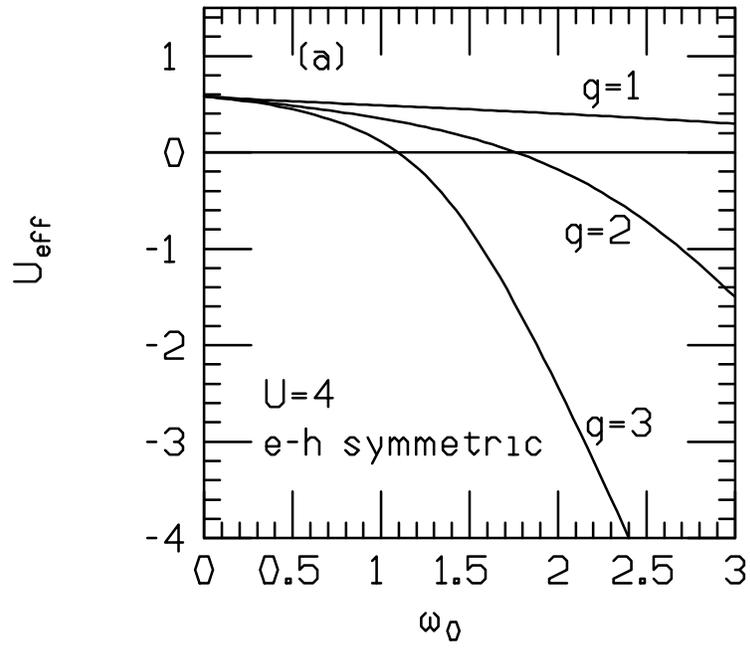
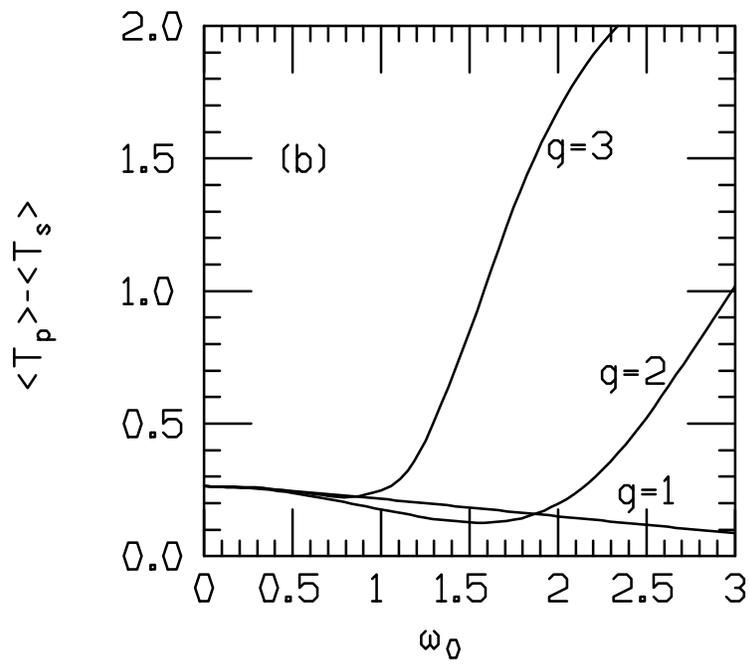
\begin{figure}
\caption { Exact diagonalization results for Holstein-like model Eq. (39). 
(a) Effective interaction and (b) difference between kinetic energy of a pair and kinetic energy
of two holes in the 4-site chain versus frequency $\omega_0$ for $U=4$ and
various values of $g$. The kinetic energy increases rapidly as the
effective interaction becomes more attractive, in contrast to the 
behavior found in the dynamic Hubbard model.
}
\label{Fig. 15}
\end{figure}

\begin{figure}
\caption { Snapshots of Monte Carlo configurations for the Holstein-like model
for an $N=20$-site
lattice with $L=40$ time slices. $\Delta \tau=0.25$, $g=3$, $\omega_0=2$, $U=4$.
Same conventions as in Fig. 9.
After a few initial sweeps the holes are still separate and light (upper panels),
after several more  sweeps (lower panels)  the holes are bound and
heavier (world lines fluctuate less in time direction), in contrast to the behavior
seen in Fig. 9.
}
\label{Fig. 16}
\end{figure}

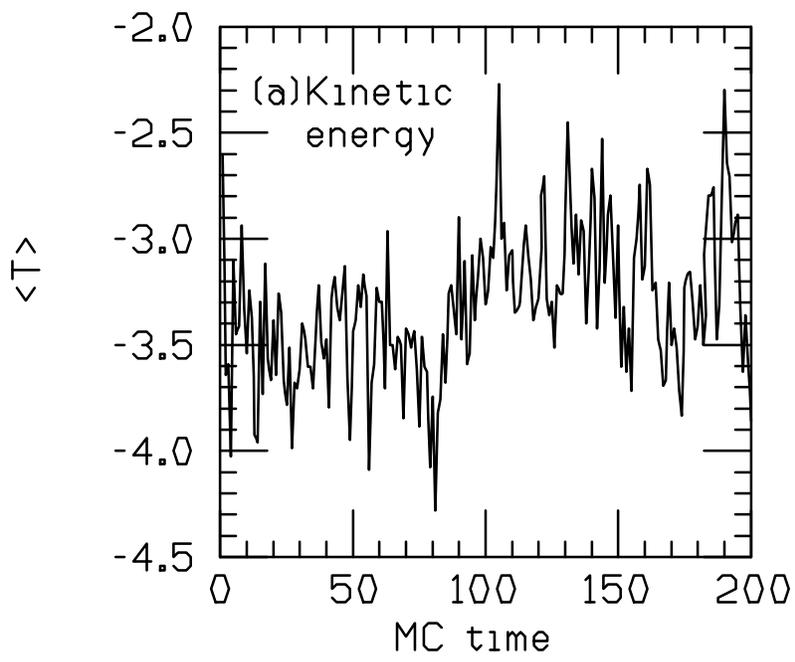
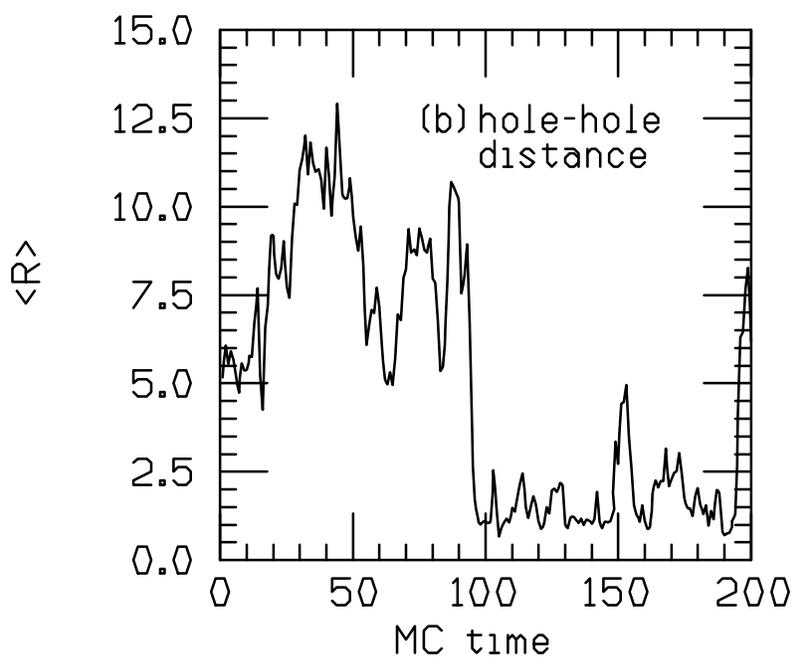
\begin{figure}
\caption {Kinetic energy of two holes (a) and hole-hole distance (b) as function
of 'Monte Carlo time' for the Holstein-like model. The unit of time is 15 Monte Carlo sweeps.  
$N=20$, $L=40$, $\Delta \tau=0.25$, $g=3$, $\omega_0=1$, $U=3$. After about 100
steps the hole-hole distance decreases drastically as the holes become bound (b),
and at the same time the kinetic energy increases (a), in contrast to the behavior seen
in Fig. 10.
}
\label{Fig. 17}
\end{figure}

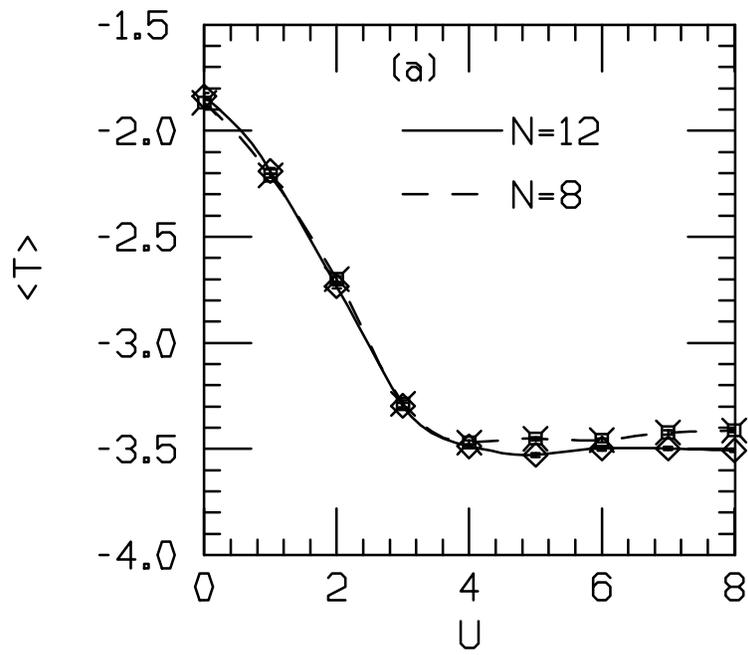
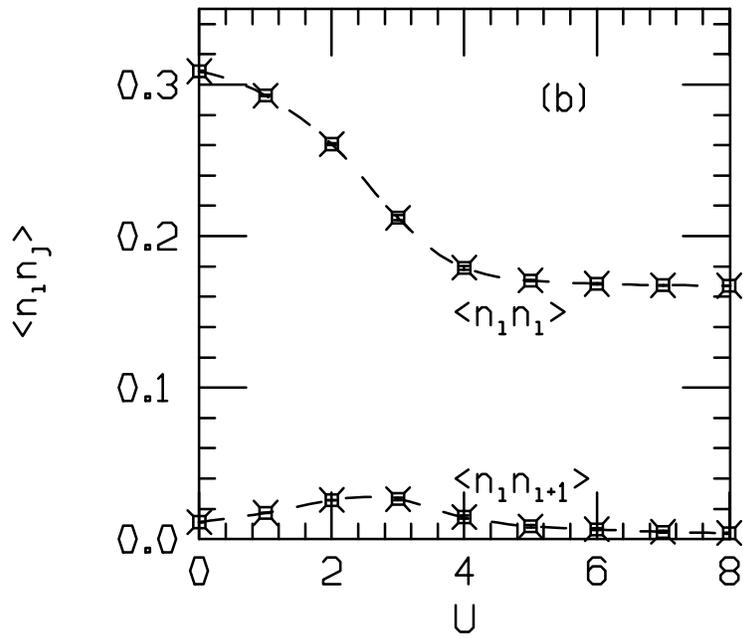
\begin{figure}
\caption { (a) Kinetic energy and (b) on-site and nearest neighbor density-density 
correlation versus on-site repulsion $U$ from Monte Carlo
simulations of the Holstein-like model with $N=12$. In (a) results for $N=8$ are
also shown, for (b) the size dependence is negligible. As $U$ decreases and pairs
form, the on-site correlations increase and the kinetic energy increases, in contrast
to the behavior seen in Figs. 12 and 13 for the dynamic Hubbard model.
}
\label{Fig. 18}
\end{figure}

 \end{document}